\documentclass[11pt]{cernrep}
\usepackage{graphicx}
\usepackage{here}

\def\g{\gamma}

\def\ran{\rangle}

\def\ket#1{|#1\ran}

\def\e{\epsilon}
\def\frac#1,#2{{#1\over #2}}

\def\p{\vec p}

\def\fr#1,#2{{#1\over #2}}
\def\p{\bot}
\def\pa{\scriptscriptstyle \|}

\def\dd{\partial}

\def\e{\varepsilon}

\def\ls{\left(}

\def\rs{\right)}

\def\g{\gamma}

\def\n{\nu}
\def\m{\mu}

\def\La{\Lambda}

\def\ps{\psi}

\begin{document}
\begin{flushright}
SMUHEP/03-12
\end{flushright}

 \title{A Nonperturbative Calculation of the Electron's Magnetic Moment}
\author{G. McCartor}
\institute{SMU}
\maketitle
\begin{abstract}
I shall present nonperturbative calculations of the electron's magnetic moment using the light-cone representation.
\end{abstract}

\section{INTRODUCTION}

For some time we have been studying the problem of performing nonperturbative calculations in quantum field theory~\cite{bhm1,bhm2,bhm3,bhm4,bhm5,Paston:1997hs,Paston:2000fq}.  The most serious problem is to find a method of regularizing the calculation in such a way as to preserve the sacred symmetries (Lorentz and gauge invariance), or at least preserve them well enough to allow an effective renormalization to be preformed.  Since the method of regularization must also allow for efficient calculations to be performed, the problem presents a challenge.  We have been interested in the possibility of including Pauli--Villars fields in the calculation --- that is, including them in the lagrangian right from the beginning.  This preserves Lorentz invariance; in some cases it also effectively preserves gauge invariance; in those cases where it breaks gauge invariance we shall have to add counter terms. These procedures will, in some cases, give a finite theory which preserves the symmetries. But since we do not know how to find an exact solution, we must also manage to produce an approximate solution.  All the ways that we know of to do that involve truncating the Fock space.  That truncation will certainly break all of the symmetries.  We argue, though, that in the last step (the truncation) it is more a question of accuracy than symmetry.  With the regulators in place we presume that there is an exact solution which preserves the symmetries, and if our approximate solution is close to the exact one, even if the small difference is in such a direction as to maximally violate the symmetries, it is still a small difference.  Of course, the negative metric fields will violate unitarity.  We shall have more to say about that below.  For reasons also discussed below, we have chosen to formulate the method in the light-cone representation.

To further test the methods on a realistic problem to which we know the answer, we have chosen to do a nonperturbative calculation of the electron's magnetic moment.  In some ways the problem is not ideal:  the physical electron is a very perturbative object while it is in some ways ill suited to our methods; therefore we do not expect to do better, or even as good as perturbation theory.  But that is not our objective; we simply want to verify that an approximate nonperturbative solution for the electron's magnetic moment is an approximation to QED.  For that study, the problem has the advantages that we know the answer, we know that the answer is given by QED and we know the details of how the answer is given by QED.  If the method gives a useful answer to this problem, even if it is not as good as perturbation theory, we may hope that the method will give a useful answer to problems that perturbation theory cannot attempt, such as hadron bound states.

We find that there are three problems which must be solved in order to produce a useful calculation of the electron's magnetic moment: the problem of maintaining gauge invariance; the problem of uncancelled divergences; and the problem of new singularities.  We believe that we have found effective solutions to these problems, at least for the present calculations.  I shall present these solutions in later sections but first I want to present a calculation which does not work.  The justification for presenting an unsuccessful calculation is that it is the application of the standard light-cone methods to the problem of the magnetic moment.  There are important points to be found in the failure of that calculation and  that failure shows that the later successes are not entirely trivial.

\section{TROUBLE IN LIGHT-CONE GAUGE}

We use the standard $P^-$ in light-cone quantized, light-cone gauge except for the modifications due to the inclusion of the Pauli--Villars fields; it has been written down without Pauli--Villars several times~\cite{tbp}.  The problem was considered in a perturbative context by Langnau and Burkardt~\cite{lb}.  We should remark, however, that with the inclusion of any number of Pauli--Villars fermi fields, the four point interactions which would take a state of one electron and one photon to another state of one electron and one photon are missing from $P^-$; that such terms are not included below is not an omission; the calculation is complete in our chosen subspace.  We truncate the Fock space to the one electron sector plus the one electron, one photon sector.  We then solve the eigenvalue problem:
\begin{equation}
    P^- \ket{s} = M^2 \ket{s}
\end{equation}
Once we have the wave function we calculate the magnetic moment using the method of ~\cite{Brodsky:1980zm}.

We have carried through the proposed manipulations but they do not lead to a successful calculation.  The problem is that our estimate of the anomalous magnetic moment has a very strong dependence on the Pauli-Villars mass scale.  If we use units of ${\alpha \over 2 \pi}$, so that the correct value is near 1, then we find that, even with a value for the photon mass as large as 0.5 electron masses, when the Pauli-Villars mass scale changes from 3 times the electron mass to 7 times the electron mass the anomalous moment changes from 1.2 to -1.2.  If we use a smaller value for the photon mass, which we would surely have to do to get useful results, the dependence is even stronger.  Since we cannot hope to estimate the optimum value for the Pauli--Villars mass scale even to within this range, the present calculation is clearly useless.  Furthermore, the problem is clearly the loss of gauge invariance; gauge invariance should prevent such strong behavior.

We note that if we keep only the physical field and set $M = m_0 =m$, the function which appears in our eigenvalue equation is just the (unregulated) one-loop fermion self-energy:
\begin{equation}
     {e^2 \over 16 \pi^2} \int dx dz  {1 \over x} {{1 + x^2 \over (1 - x)^2} z + m^2 (1-x)^2  \over m^2 x (1-x) - m^2 (1-x) - \mu^2 x - z} \label{lca}
\end{equation}
Therefore, a very useful point of comparison is the paper of Brodsky, Roskies and Suaya~\cite{brs}.  They evaluated all the graphs needed to calculate the electron's magnetic moment in perturbation theory through order $\alpha^2$.  Included in their calculations is the one-loop electron self-energy.  They did not use light-cone quantization but wrote down time-ordered perturbation theory in the equal-time representation then boosted to the infinite momentum frame.  They worked in Feynman gauge but the electron self-energy should be gauge invariant.  They get
\begin{equation}
     {e^2 \over 16 \pi^2} \int dx dz {1\over x}{z + m^2(1 - 4 x - x^2) \over m^2 x (1-x) - m^2 (1-x) - \mu^2 x - z}
\end{equation}
While the integrands do not have to be identical, the results of the integrations should be equal.  Whether we regulate with Pauli-Villars electrons, Pauli-Villars photons or a combination of both, the integrals are not equal.  If we tried to use a momentum cutoff the results would be even worse.

It is clear that the problem is that gauge invariance has been lost in solving the constraint equation
\begin{equation}
       \partial_-^2 A^- + \partial_-\partial_i A^i = -e \Psi_+^\dagger \Psi_+
\end{equation}
We do not yet know exactly what has gone wrong with solving the equation.  It is possible that the wrong boundary conditions have been used or that the equation is wrong as it stands: it must be the equation satisfied by the regulated fields and something like Schwinger terms might need to be included.  We hope to report further on the details of a resolution to the problem of the loss of gauge invariance in what may be thought of as standard light-cone techniques, but we will not consider the problem further in the present paper. Instead, we will turn our attention to ways that we know of to fix the problem.  These ways involve the use of other gauges or other methods of regulation.

\section{UNCANCELLED DIVERGENCES}

The problem of uncancelled divergences was not discovered by us but has been known for a long time.  It occurs anytime we truncate the Fock space.  It is perhaps easiest to understand by comparing with perturbation theory, for example, for the electron' magnetic moment considered in this paper.  If we truncate the space to include only the subspace of one electron and the subspace of one electron and one photon, calculate the wave function nonperturbatively and use that wave function to calculate the magnetic moment, we get a result of the form:
\begin{equation} 
\kappa = {g^2 [finite quantity] \over 1 + g^2[finite quantity] + g^2[finite quantity] \log \mu_2} 
\end{equation}
where $\mu_2$ is the Pauli--Villars mass scale.  If we let $\mu_2$ go to infinity without allowing the coupling constant to vary, we will get zero.  That would not happen in perturbation theory: since the numerator is already order $g^2$ we would use only the 1 from the denominator and might get a nonzero result.  In order $g^4$ we would use the divergent term in the denominator but there would be new terms in the numerator which would contain cancelling divergences.  This is the problem of uncancelled divergences. While we can find ways of allowing the bare mass and the coupling constant to depend on the Pauli--Villars masses that will allow a finite result to be obtained, the results we obtain do not look anything like the results from perturbation theory and generally do not make much sense physically.  The solution is to keep the Pauli--Villars masses finite.  We think of it this way: If the limit of infinite Pauli--Villars masses would give a useful answer in the case where we do not truncate the Fock space (so, we have no uncancelled divergences), then there must be some finite value of the Pauli-Villars masses that would also give a useful answer.  The question is whether we can use a sufficiently large value.  To answer that question we must consider that there are two types of error associated with the values of the Pauli--Villars masses.  The first type of error results from having these masses too small; then our wave function will contain too much of the negative normed states, unitarity will be badly violated and in the worst case we might get negative probabilities.  That type of error goes like
\begin{equation}
    E_1 \sim {M_1/M_2}
\end{equation}
Where $M_1$ is the physical mass scale and $M_2$ is the Pauli--Villars mass scale.  The other type of error results from having the Pauli--Villars masses too large; in that case our wave function will project significantly onto the parts of the representation space excluded by the truncation.  That error can be roughly estimated as
\begin{equation}
    E_2 \sim {\langle \Phi_{+}^\prime|\Phi_{+}^\prime\rangle \over \langle \Phi_{+}|\Phi_{+}\rangle}
\end{equation}
where $|\Phi_{+ }^\prime\rangle$ is the projection of the wave function onto the excluded sectors.  In practice, this quantity can be estimated by doing a perturbative calculation using the projection onto the first excluded Fock sector as the perturbation.  If both types of error are small, we can do a useful calculation; otherwise not.

The main reason for thinking that we might be able to do a useful calculation in spite of the problem of uncancelled divergences is the lesson from earlier studies mentioned above: the rapid drop off of the projection of the wave function onto higher Fock sectors.  Just where this rapid drop off occurs depends on the theory, the coupling constant and the values of the Pauli--Villars masses.  At weak coupling and relatively light Pauli--Villars masses only the lowest Fock sectors are significantly populated.  At stronger coupling or heavier Pauli--Villars masses, more Fock sectors will be populated; but eventually the projection onto higher sectors will fall rapidly.  The rapid drop off of the projection of the wave function onto sufficiently high Fock sectors is the most important reason why we do our calculations in the light-cone representation.  For any practical calculations on realistic theories we have to truncate the space and we must have a framework in which that procedure can lead to a useful calculation.  The rapid drop off in the projection of the wave function will not happen in the equal-time representation mostly due to the complexity of the vacuum in that representation.  These features can be explicitly demonstrated by setting the Pauli--Villars masses equal to the physical masses.  In that case the theory becomes exactly solvable~\cite{bhm4}.  The spectrum is the free spectrum and the theory is not useful for describing real physical processes due to the strong presence of the negative normed states in physical wave functions, but it still illustrates the points we have been trying to make.  In that theory the physical vacuum is the bare light-cone vacuum while it is a very complicated state in the equal-time representation.  Physical wave functions project onto a finite number of Fock sectors in the light-cone representation but onto an infinite number of sectors in the equal-time representation.  While the operators that create the physical eigenstates from the vacuum are more complicated in the equal-time representation than in the light-cone representation, the major source of the enormous complication of the equal-time wave functions is the equal-time vacuum.  As the Pauli--Villars masses become larger than the physical masses, the light-cone wave functions project on to more of the representation space and more so as the coupling constant is larger and the Pauli--Villars masses are larger, but the wave functions remain much simpler than in the equal time representation and to the extent we can do the calculations, there is always a point of rapid drop off of the projection onto higher Fock sectors.  Due to the rapid drop off of the projection of the wave function onto the higher Fock sectors we believe that, given a value of the Pauli--Villars mass scale sufficiently large to assure that the first type of error discussed above --- the one due to taking the Pauli--Villars masses too small--- is small, we could find a sufficiently large part of the representation space such that the second type of error will also be small.  Therefore, we do not believe that the need to keep the value of the Pauli--Villars masses finite imposes any fundamental limitation on the accuracy which could, in principle, be achieved.  Of course, even if that is true it is not guaranteed that the required part of the representation space would be small enough to allow us to perform accurate calculations.  Furthermore, for the method to be useful in practice, there must not only be a value of the Pauli--Villars mass for which both types of errors are small but there must be a wide range of such values since the optimum value for the Pauli--Villars mass can only be rather crudely estimated.  A principal objective of the present work is to test these ideas on a physically realistic problem to which we know the answer.

\section{NEW SINGULARITIES}

To perform the nonperturbative calculations we must face the problem of new singularities.  We have to do integrals with denominators of the form $(- M^2 x (1-x) + m^2 x + \mu^2 (1-x) + z)$, where $M$ is the physical electron mass, $m$ is the bare electron mass and $\mu$ is the photon mass.  When the bare mass is less than the physical mass, as is the case in QED, there can be a zero in this denominator.  In perturbation theory the expansion is about $M = m$ and the denominator cannot vanish as long as the photon is given a small nonzero mass (or a large nonzero mass if it is a Pauli--Villars photon).  The standard techniques in perturbation theory thus avoid this singularity.  We find that when the zero is a simple pole, the principle value prescription is correct.  But in the wave function normalization the denominator is squared so there is a double pole and we must give it a meaning.  We believe that the correct prescription is
\begin{eqnarray}
   &&\int dy \; dz \;  y \;{f(y,z)\over [ m^2 y + \mu_0^2 (1-y) -M^2 y (1-y) + z]^2}  \equiv  
 \lim_{\epsilon\rightarrow 0} \nonumber \\
&&{1\over 2 \epsilon} \int dy \int dz f(y,z)
\Bigg[{1 \over [ m^2 y + \mu_0^2 (1-y) - y (1-y) + z - \epsilon]}\nonumber \\
&&- {1 \over [ m^2 y + \mu_0^2 (1-y) - M^2 y (1-y) + z + \epsilon]}\Bigg]
\end{eqnarray}
This prescription has the interesting consequence that the wave function normalization is infrared finite whereas it is infrared divergent in perturbation theory.  The reason is that, with the prescription, the true singularity occurs at $M = m + \mu$; in perturbation theory, with $M = m$, this is at $\mu = 0$, which is the infrared singularity and the reason that the photon mass cannot be taken all the way to zero in perturbation theory.  For the nonperturbative calculation, the physical photon mass can be taken to zero since $M \neq m$.  The basic requirement of these prescriptions is that they preserve the Ward identities.  We have not shown that the prescription preserves the Ward identities but the answers we get do depend on the prescription and we believe that it passes the test so far.

\section{FEYNMAN GAUGE}

In this section we shall calculate the electron's magnetic moment using Feynman gauge.  We shall regulate the theory by the inclusion of one Pauli-Villars photon and one Pauli-Villars fermion with the inclusion of flavor changing currents.  The Lagrangian is thus
\begin{equation}
  \sum_{i=0}^1 -{1 \over 4} (-1)^i F_i^{\mu \nu} F_{i,\mu \nu} -{\lambda_i \over 2} (\partial_\mu A_i^\mu)^2 + \sum_{i=0}^1 (-1)^i \bar{\psi_i} (i \gamma^\mu \partial_\mu - m_i) \psi_i - e \bar{\psi}\gamma^\mu \psi A_\mu 
\end{equation}
where
\begin{equation}
  A^\mu  = \sum_{i=0}^1 A^\mu_i \quad \psi = \sum_{i=0}^1 \psi_i \quad F_i^{\mu \nu} = \partial^\mu A_{i}^{\nu}-\partial_\nu A_{i}^{\mu}
\end{equation}
Here, $i=0$ are the physical fields and $i=1$ are the Pauli-Villars (negative metric) fields.

We must remark on two effects of including the Pauli-Villars fermi fields with the flavor changing currents, one good effect and one apparently bad effect.  The first effect, the good one, pertains to the operator $P^-$.  If one works out $P^-$ including only the physical fields one encounters the need to invert the covariant derivative $\partial_- - e A_-$ ~\cite{sb}.  The same problem occurs in any gauge where $A_-$ is not zero.  That complication is perhaps the main reason that gauges other than light-cone gauge have received relatively little attention in the light-cone representation.  While the inverse of the covariant derivative can be defined by a power series in $e$, or, in a truncated space may be calculated exactly if the truncation is sufficiently severe, one has the feeling that $P^-$ has not been fully written down since there remains the nontrivial problem of inverting the covariant derivative.  But with the inclusion of the Pauli-Villars fermions with the flavor changing currents we find that the problem does not occur: the inverse of the covariant derivative is replaced by the inverse of the ordinary derivative.  The second effect of the flavor changing currents (the apparently bad effect) is that they break gauge invariance.  That would seem to require that we include counter terms in the lagrangian to correct for the breaking of gauge invariance.  It turns out that that is not necessary.  The reason is that we can take the limit of the Pauli-Villars fermion mass, $m_2$, going to infinity.  One might properly worry that there might still be finite effects of the necessary counter terms but the counter terms go to zero like powers of $m_2$ while the only divergences we encounter are logs.  We shall therefore proceed with the calculation using only the Lagrangian given above.

We expand the wave function as:
\begin{equation}
  |\psi\rangle = b_{0,+}(1,\vec{0}) |0\rangle + \sum_{s,\mu,i,j} \int C^\mu_{s,i,j} b_{is}^\dagger(x,\vec{k}) a^{\mu \dagger}_j(1-x,\vec{k}) |0\rangle
\end{equation}
where we have set the total +-momentum of the state to 1 and the total transverse momentum of the state to zero; we shall also take the physical mass of the electron to be 1.  In principal, we should include a term which is a state of one Pauli-Villars fermion; but we find that if we do, the coefficient of that term goes to zero when $m_2$ goes to infinity (see also~\cite{bhm4})so we will not include it here.

We find that the eigenvalue equation takes the form
\begin{equation}
  1 - m_0^2 = 2 G  \int dx\; dz \sum_{i,j}{(-1)^{i + j} \over x}\left[{(m_j^2 - 4 m_0 m_j x + m_0^2 x^2) -z \over  x (1-x) - m_j^2 (1-x) - \mu_i^2 x  - z}\right]
\end{equation}
One indication that we have successfully implemented gauge invariance is that this integral is just that of ref.~\cite{brs} with the perturbative denominator replaced by the nonperturbative denominator.  When we complete the calculation of the wave function and then calculate the magnetic moment using the method of ~\cite{Brodsky:1980zm}, we find that, in units of the Schwinger term, ${\alpha \over 2 \pi}$, the anomalous moment, $\kappa$ is given by: $\kappa = .99$ at $\mu_1 = 3$, $\kappa =1.09$ at $\mu_1 = 10$, $\kappa = 1.13$ at $\mu_1 = 100$ and $\kappa = 1.49$ at $\mu_1 = 1000$.  Thus, we find that the magnetic moment of the electron,$\mu$ is probably between 1.0011 $\mu_0$ and 1.0013 $\mu_0$ where $\mu_0$ is the Dirac moment.  If we estimate the optimum Pauli-Villars  mass scale using the method given above we find that it is about 30 times the electron mass and obtain an estimate of about 1.0012 $\mu_0$.  In the absence of the more accurate perturbative estimate that would be a useful calculation to compare with experiment and we consider the outcome to be entirely satisfactory.

\section{LIGHT-CONE GAUGE AGAIN}

Although we did not succeed in our earlier light-cone gauge calculation based on regulating the standard light-come method with Pauli-Villars fields, we can do a successful light-cone gauge calculation based on the method of~\cite{Paston:2000fq}.  In that reference the authors show that the regularization method gives perturbative equivalence with standard Feynman methods.  The present calculations are the first use of the method in a nonperturbative calculation.  The Lagrangian is

\begin{eqnarray}
&&L=-\frac{1},{4}\sum_{j=0,1}(-1)^jF_j^{\m\n}
\ls 1+\frac{\dd_{\pa}^2},{\La^2_j}-
\frac{\dd_\p^2},{\La^2}\rs F_{j,\m\n} \\
&&+\sum_{l=0}^{3}\frac{1},{v_l}\bar\ps_l\ls i\g^\m\dd_\m-M_l\rs \ps_l
-e A_\m \bar\ps \g^\m\ps
\end{eqnarray}
where

\begin{equation}
F_{j,\m\n}=\dd_\m A_{j,\n}-\dd_\n A_{j,\m},\quad
v_0=1,\quad \sum_{l=0}^3 v_l=0,\quad \sum_{l=0}^3 v_lM_l=0,\quad 
\sum_{l=0}^3 v_lM_l^2=0
\end{equation}
\begin{eqnarray}
A_\m=A_{0,\m}+A_{1,\m},\quad \ps=\sum_{l=0}^3\ps_l,\quad
\frac{1},{\La_j^2}=\cases{1/\La^2,
&$j=0$,\cr 1/\La^2+1/\m^2, &$j=1$\cr}
\end{eqnarray}
There are two other regulation parameters which remove certain states form the Fock space: $\e$ ($\e>0$ cuts off the small $p_-$states according to, $|p_-|\ge\e$), and
$v$ ($v>0$  cuts off the small $p_\p$values according to , $|p_\p|\ge v$).
The regularization parameters are removed in a strict order according to:
 $\e\to 0$, then $\mu \to 0$, then $v \to 0$, then $M_L \to \infty$.  All these limits are finite and after taking them we are left with a theory regulated only by $\La$ which acts something like a photon mass.  As seen in the Lagrangian, it is not exactly a photon mass but rather a parameter which controls the higher derivatives in the photon kinetic energy.  In the final answer $\La$ appears in much the same way as the Pauli-Villars photon mass does in Feynman gauge.

We have not quite completed the calculations in light-cone gauge using the more sophisticated regularization method but the calculations are nearly done and we can say that the final answers will be very close to those in Feynman gauge.

\end{document}